\documentclass
[aip,reprint,showpacs,floats,amsmath,amssymb,a4paper,10pt,tightenlines,twocolumn]{revtex4-1}%
\usepackage{epsf}
\usepackage{graphicx}
\usepackage{amsmath}
\usepackage{amsfonts}
\usepackage{amssymb}
\usepackage{dcolumn}
\usepackage{doublespace}
\usepackage{layout}%
\providecommand{\U}[1]{\protect\rule{.1in}{.1in}}

\begin{document}
\title{Spin torque transistor revisited}
\date{\today}
\author{Takahiro Chiba}
\email{t.chiba@imr.tohoku.ac.jp}
\affiliation{Institute for Materials Research, Tohoku University, Sendai 980-8577, Japan}
\author{Gerrit E. W. Bauer}
\affiliation{Institute for Materials Research, Tohoku University, Sendai 980-8577, Japan}
\affiliation{WPI-AIMR, Tohoku University, Sendai 980-8577, Japan}
\affiliation{Kavli Institute of NanoScience, Delft University of Technology, Lorentzweg 1,
2628 CJ Delft, The Netherlands}
\author{Saburo Takahashi}
\affiliation{Institute for Materials Research, Tohoku University, Sendai 980-8577, Japan}

\begin{abstract}
We theoretically study the operation of a 4-terminal device consisting of two
lateral thin-film spin valves that are coupled by a magnetic insulator such as
yttrium iron garnet (YIG) via the spin transfer torque. By magnetoelectronic
circuit theory we calculate the current voltage characteristics and find
negative differential resistance and differential gain in a large region of
parameter space. We demonstrate that functionality is preserved when the
control spin valve is replaced by a normal metal film with a large spin Hall angle.

\end{abstract}
\maketitle


A transistor is a three terminal device that plays important roles in today's
electronics. A conventional transistor generates a large current modulation
between source and drain terminals by a relatively small signal on the third
\textquotedblleft base\textquotedblright\ contact. This property is called
\textquotedblleft gain\textquotedblright\ and the corresponding circuit acts
as an \textquotedblleft amplifier\textquotedblright. In the field of
spintronics three-terminal devices have been studied since Datta-Das proposed
the spin FET,\cite{Datta90} in which the electronic spin degrees of freedom
are utilized to achieve new functionalities in circuits and devices made of
ferromagnetic and normal conductors. However, with few
exceptions\cite{Bauer03,Konishi12} spin transistors lack current gain, which
is essential for many applications.

A transistor based on the current-induced spin-transfer torque, the so-called
spin torque transistor (STT), was proposed a decade ago.\cite{Bauer03} Figure~1
shows the schematics of the device. The central insulating ferromagnetic disk
with in-plane magnetization is sandwiched by normal metal films on both sides
that form the spacers of two lateral spin valves (LSVs). The magnetizations in
the upper and lower LSVs are parallel to the $x$ and $y$ direction,
respectively, forming a closed magnetic flux loop with weak stray fields. An
applied voltage $V_{S}$ drives a current through the lower LSV, generating a
spin accumulation in the lower normal metal spacer that exerts a torque on the
magnetization of the central magnetic disk in the $y$-direction. Application
of a voltage $V_{B}$ induces a spin accumulation that creates a spin transfer
torque along $x$, which competes with that of the lower LSV. The magnetization
direction of the central layer can therefore be controlled by the relative
magnitude of $V_{S}$ and $V_{B}$. The transistor action consists of the
control of the source-drain current $I_{SD}$ by the base voltage $V_{B}$. This
device can display negative differential resistance and gain when the
conductance polarization is high and spin-flip scattering is small, even at
room temperature.\cite{Bauer03} Unfortunately, current gain was found only for
very highly polarized magnetic contacts. The originally proposed structure was
also complicated, since the central layer was assumed to be a strongly coupled
magnetic tunnel junction.

\begin{figure}[ptb]
\begin{center}
\includegraphics[width=0.4\textwidth,angle=0]{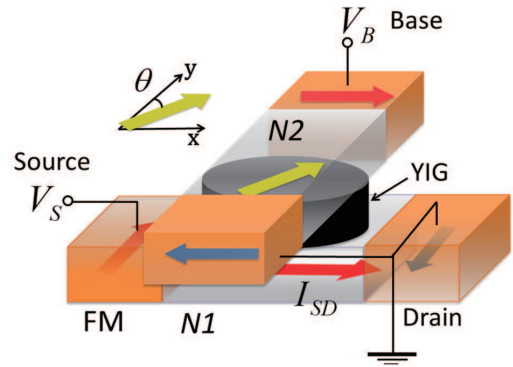}
\end{center}
\caption{Schematics of the spin torque transistor.\cite{Bauer03} The contacts
are ferromagnets in a flux closure configuration. The circular disk is made
form a magnetic insulator with easy plane magnetization, while the rectangles
represent normal metal thin films. $\theta$ is the angle between base
magnetization and $y$-axis.}%
\label{Figure1}%
\end{figure}

Recently, magnetic insulators have attracted attention as new materials for
spintronics. The magnetization of Yttrium iron garnet (YIG), a ferrimagnetic
insulator (FI) with a large band gap, can be activated
thermally\cite{Uchida10} or electrically\cite{Kajiwara10} by means of the spin
Hall effect (SHE) via a Pt contact and detected electrically in another Pt
contact using the inverse SHE (ISHE). Spin transport at a normal metal
(N)\textrm{%
$\vert$%
}FI interface is governed by the spin-mixing conductance $g^{\uparrow
\downarrow}$.\cite{Brataas00} The prediction of large $g^{\uparrow\downarrow}$
for interfaces between YIG and simple metals by first-principle
calculations\cite{Jia11} has been confirmed by experiments,\cite{Heinrich11}
proving that the magnetization in insulators may undergo large spin-transfer
torques. We therefore propose here a thin YIG film as central layer of an STT
as shown in Fig. 1. Secondly, we suggest to take advantage of the recent
discovery of the giant spin Hall effect in Ta\cite{Tant} and W\cite{Tung} or
Bi-doped Cu\cite{CuBi} to operate the STT, leading to further simplifications
of the device design.

The $I$-$V$ characteristics of the spin torque transistor with a YIG base as shown
in Fig. 1 can be computed by magnetoelectronic circuit theory.\cite{Brataas00}
We calculate source-drain currents, torques on the base magnetization created
by the spin accumulations, and the differential resistance and gain as a
function of the voltage ratio $V_{S}/V_{B}$ and device parameters.

At the interface between a monodomain ferromagnet with magnetization parallel
to the unit vector $\mathbf{m}$ and a paramagnetic metal, the charge and spin
currents, $I_{c}$ and $\boldsymbol{I}_{s}$, driven by charge chemical
potential difference $\Delta\mu_{c}$ and spin accumulation in the normal metal
$\Delta\boldsymbol{\mu}_{s}$ are linear functions of the interface
conductances. The conventional conductances $g^{\uparrow\uparrow}$ and
$g^{\downarrow\downarrow}$ for electrons with up and down spins, respectively,
vanish when the ferromagnet is an insulator. The complex spin-mixing
conductance $g^{\uparrow\downarrow}$ governs the spin current polarized
transverse to the magnetization. The conductance parameters are in units of
the conductance quantum $e^{2}/h$, contain (for ferromagnetic metals) bulk and
interface contributions, and can be computed from
first-principles.\cite{Xia10} For metallic\cite{Xia10,Stiles12} and
insulating\cite{Jia11} ferromagnets, $\operatorname{Im}g^{\uparrow\downarrow}$
is usually smaller than $10\%$ of $\operatorname{Re}g^{\uparrow\downarrow}$
and is disregarded below. It is convenient to introduce $g=g^{\uparrow
\uparrow}+g^{\downarrow\downarrow}$ and $p=(g^{\uparrow\uparrow}%
-g^{\downarrow\downarrow})/g$, where $g$ is the total conductance and $p$ its
polarization. The continuity equation for spin current and spin accumulation
$\boldsymbol{\mu}_{s}^{N1}$ in N1 reads:
\begin{equation}
\mathbf{I}_{s}^{S}+\mathbf{I}_{s}^{D}+\mathbf{I}_{s}^{B}=\frac{e^{2}%
N(0)V_{ol}}{\tau_{\mathrm{sf}}}\boldsymbol{\mu}_{s}^{N1},
\end{equation}
where $\boldsymbol{I}_{s}^{S/D/B}$ are the spin currents flowing from the
Source/Drain/Base ferromagnets into the spacer N1. $N(0)$ and $V_{ol}$ are the
density of states at the Fermi level and the volume, and $\tau_{\mathrm{sf}}$ is the
spin-flip relaxation time. Spin-flip can be disregarded in the normal metal
node of small enough structures made from metals with weak spin dissipation
such Al,\cite{Jedema00} Cu,\cite{Jedema01} Ag,\cite{Silver} or graphene.\cite{graphene} 
The spin-flip in the source and drain electrodes can simply be included by taking
their magnetically active thickness as the smaller one of the spin-flip
diffusion length and physical thickness. The electrically insulating base
electrode is assumed to be thin and magnetically soft. The source-drain
current $I_{SD}$ has been derived earlier\cite{Bauer10} in terms of $g_{S}$
and $p_{S}$, the normal conductance and polarization of the metallic
source/drain contacts and $g_{S}^{\uparrow\downarrow}$ $\bigl(g_{B}%
^{\uparrow\downarrow}\bigl)  $ is the spin-mixing conductance of the
source/drain (insulating base) contacts. $I_{SD}$ depends on the base
magnetization angle $\theta$ with respect to the $y$-axis. The torque
$\tau_{B}^{N1}(\theta)$ on the base magnetization created by the spin
accumulation in the space is proportional to the transverse spin current into
the base.\cite{Bauer10} We disregard effects of the \O rsted field produced by
$I_{SD}.$ A steady state with finite $\theta$ exists when $\tau_{B}^{N1}$ is
exactly canceled by an external torque, either from an applied magnetic field
or a current-induced torque from the top layer. We assume the same parameters
for the upper and lower sections such that $\tau_{B}^{N2}(\theta)/V_{B}=\tau_{B}%
^{N1}(\pi/2-\theta)/V_{S}$ (see Fig. 1), where $V_{B}$ is the voltage over the
upper layer. We keep the ratio between the mixing conductances of metal and
insulator variable, \textit{viz}. chose $g_{S}:g_{S}^{\uparrow\downarrow
}:g_{B}^{\uparrow\downarrow}=1:1:\beta$. The stationary state of the biased
spin torque transistor is described by the angle $\theta_{0}$ at which the two
torques on the base magnet cancel each other. $\tau_{B}^{N1}(\theta_{0}%
)=\tau_{B}^{N2}(\theta_{0})$ then leads to the transcendental equation
\begin{equation}
\frac{V_{S}}{V_{B}}=\frac{\tan^{2}\theta_{0}+\epsilon}{\epsilon\tan^{2}%
\theta_{0}+1}\frac{1}{\tan\theta_{0}},\label{Eqltheta}%
\end{equation}
where $\epsilon=(\beta+2)/(2\beta+2)$. With $\delta=1/\left(  \beta+1\right)
,$ the source-drain conductance becomes
\begin{equation}
\frac{I_{SD}(V_{S},V_{B})}{V_{S}}=\frac{e}{h}\frac{g_{S}}{2}\left(
1-p_{S}^{2}\frac{\epsilon+\delta\tan^{2}\theta_{0}}{\epsilon+\tan^{2}%
\theta_{0}}\right).  \label{ISDSV}%
\end{equation}
With increasing $p_{S}$, strong non-linearities develop that for large
polarizations lead to negative differential conductances for $V_{S}%
/V_{B}\gtrsim1$.

\begin{figure}[ptb]
\begin{center}
\includegraphics[width=0.4\textwidth,angle=0]{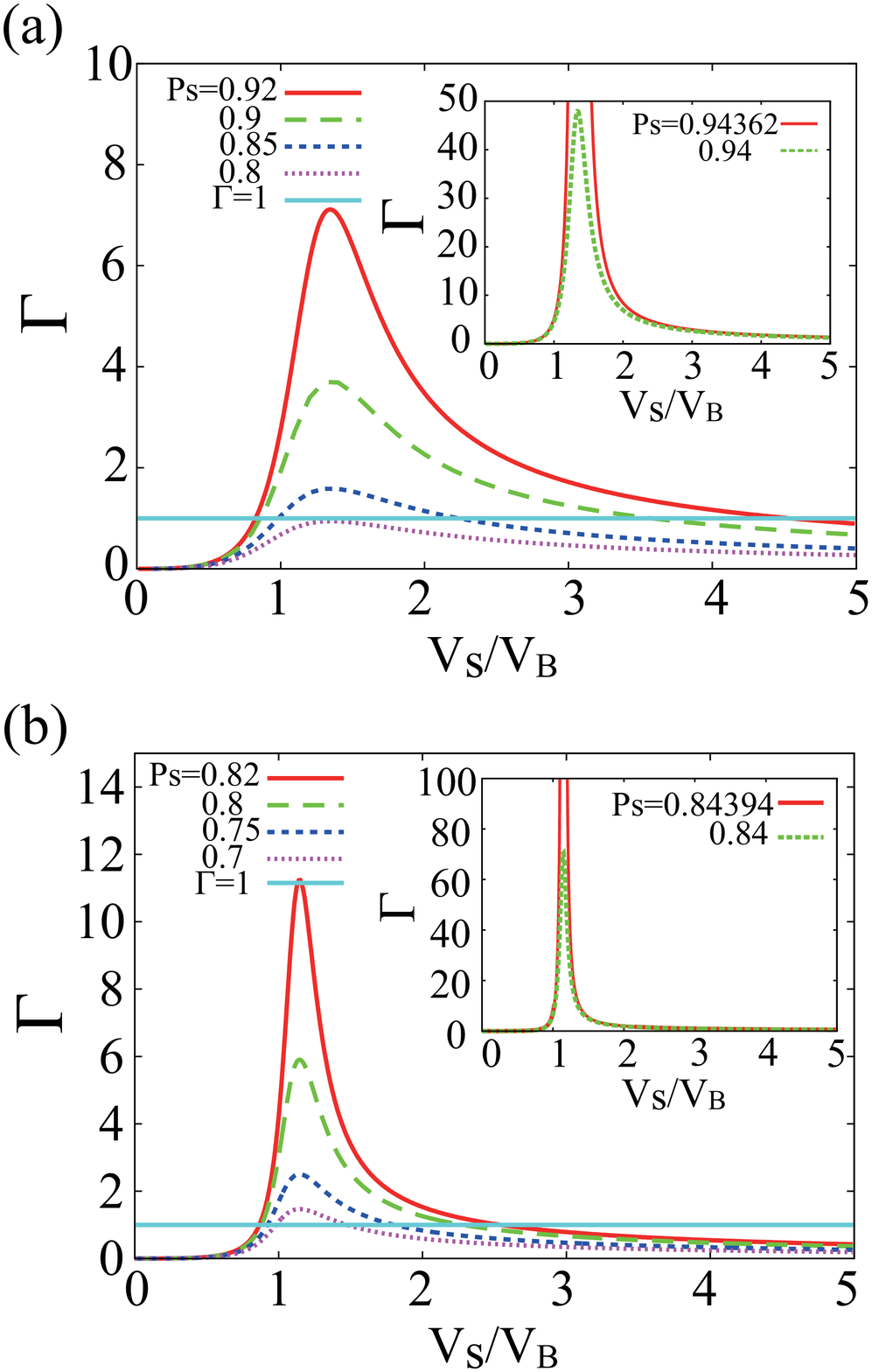}
\end{center}
\par
\caption{(a),(b):Differential current gain as a function of the voltage ratio
$V_{S}/V_{B}$ for different values of $p_{S}$ (c) $\beta=1$, (d) $\beta=5$. Insets represent the infinity gain for the critical value of $p_{S}$.}%
\label{Figure2}%
\end{figure}
\begin{figure}[ptb]
\begin{center}
\includegraphics[width=0.4\textwidth,angle=0]{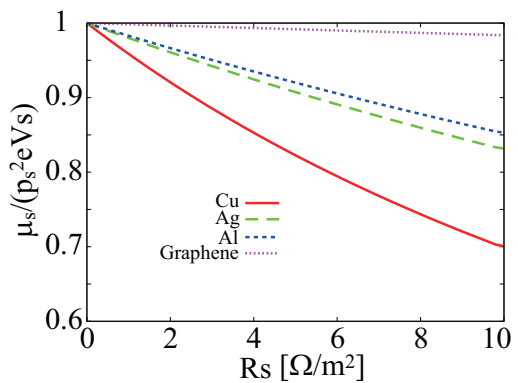}
\end{center}
\caption{Spin accumulation in the source-drain spin valve for $\theta=0$ as a
function of the interface resistance $R_{S}=hA/e^{2}/g_{S}$ for a node length
of $L_{N1}=200\,\mathrm{nm}$.}%
\label{Figure3}%
\end{figure}

We concentrate on the differential current gain $\Gamma=T/G$ as a
representative figure of merit, where $T=\left(  dI_{SD}/dV_{B}\right)
_{V_{S}}=\left(  \partial I_{SD}/\partial\theta\right)  _{V_{B}}\left(
\partial\theta/\partial V_{B}\right)  _{V_{S}}$ is the differential
transconductance and $G=\left(  dI_{SD}/dV_{S}\right)  _{V_{B}}=I_{SD}%
/V_{S}+\left(  \partial I_{SD}/\partial\theta\right)  _{V_{S}}\left(
\partial\theta/\partial V_{S}\right)  _{V_{B}}$ the differential source-drain
conductance $G$. While Ref.
\onlinecite{Bauer03}
focused on angles $\theta_{0}\rightarrow0$, we extend the calculations of the
spin torque transistor device characteristics to arbitrary working points
$\theta_{0}$ controlled by the ratio of the applied voltages. The differential
gain then reads%
\begin{widetext}
\begin{equation}
\Gamma=\frac{2p_{S}^{2}(1-\delta)\tan\theta_{0}}{1+(3\epsilon-1/\epsilon
)\tan^{2}\theta_{0}+\tan^{4}\theta_{0}-p_{S}^{2}\left[  1+(3\epsilon
-\delta/\epsilon)\tan^{2}\theta_{0}+\delta\tan^{4}\theta_{0}\right].
}\label{DG1}%
\end{equation}
\end{widetext}
By substituting the solution of Eq. (\ref{Eqltheta}), we calculate the
differential current gain $\Gamma$ as a function of $V_{S}/V_{B}\ and$ plot it
in Fig. \ref{Figure2} as a function of the ratio $V_{S}/V_{B}$ and different
values of the conductance polarization of the metallic ferromagnetic contacts
$p_{S}$. The differential current gain can be huge, particularly near the
half-metallic limit of $p_{S}=1$, indicating that the contacts should be
fabricated from high polarizations materials such as certain Heusler alloys or
very thin MgO tunnel junctions. The device performance depends strongly on all
parameters and is by no means universal. The critical value of $p_{S}$ for
vanishing differential resistance $G(\theta_{0})=0$ can be computed as
\begin{align}
p_{S}(\beta) &  =\sqrt{\frac{9\epsilon^{2}-1}{9\epsilon^{2}-\delta}}%
=\sqrt{\frac{5\beta^{2}+28\beta+32}{9\beta^{2}+32\beta+32}}\\
&  \rightarrow\frac{\sqrt{5}}{3}\simeq0.74536\ \ (\beta\rightarrow\infty).
\end{align}
$\beta$ can be increased by reducing the source/drain contact areas or by
introducing tunnel junctions, although this will increase the response time.
It should also be kept in mind that our results are valid only when the spin
accumulation is not strongly affected by spin flip assuming that $\mu_{s}%
^{N1}/p_{S}eV_{S}=1$. The error involved can be estimated by the spin
accumulation of the spin valve for $\theta=0$, for which the spin accumulation
is limited by the spin relaxation according to\cite{SA}
\begin{equation}
\frac{\mu_{s}^{N1}}{p_{S}^{2}eV_{S}}=\left[  1+\frac{hA}{e^{2}g_{S}}%
\frac{L_{N1}}{\rho_{N1}\lambda_{N1}^{2}}\right]  ^{-1},
\end{equation}
where $\rho_{N1}$ is the bulk resistivity, $L_{N1}$ the length, and
$\lambda_{N1}$ the spin diffusion length. Figure \ref{Figure3} shows this
ratio for Py$|$Cu,\ Py$|$Ag,\ Py$|$Al,\ Co$|$Graphene,\ using the (room
temperature) parameters. $\rho_{N}=2.9,\ 2.0,\ 3.2$ and 3.0\ $\mathrm{\mu
\Omega cm}$,\ and $\lambda_{N}=400,$\ 700,\ 600 and 2000\ $\mathrm{nm}$ for
Cu,\ Ag,\ Al\ and Graphene, respectively.\cite{Jedema03,Silver,Han12}%

The spin Hall effect (SHE)\cite{Dyakonov10} refers to the spin current induced
transverse to a charge current through a nonmagnetic material with spin-orbit
interaction. Recently, large spin Hall effects have been reported in
platinum,\cite{Kimura10} and CuBi alloy.\cite{CuBi} $\beta$%
-tantalum,\cite{Tant} and $\beta$-tungsten,\cite{Tung} generate spin Hall
currents large enough to induce spin-torque switching of ferromagnetic
contacts. The strength of the SHE is measured by the spin Hall angle defined
by the ratio, $\alpha_{\mathrm{SH}}=I_{s}/I_{c}$, where $I_{s}$ is the
transverse spin current induced by a charge current $I_{c}$. $\alpha
_{\mathrm{SH}}=0.07$ for Pt,\cite{Kimura10} $-0.15$ for $\beta$-Ta,\cite{Tant}
$-0.3$ for $\beta$-W,\cite{Tung} and $-0.24$ for CuBi alloy\cite{CuBi} have
been reported. We therefore suggest the device which we call the spin
Hall torque transistor. In the new device, the
control spin valve (upper one in Fig. \ref{Figure1}) is replaced by a normal metal film with a large spin Hall angle. For the cited values of $\alpha_{\mathrm{SH}}$ its
performance is comparable to the one discussed above, but easier to fabricate.
We point out the interest of simple bilayers of a spin Hall metal and magnetic
insulator, in which an new effect has been discovered recently, \textit{viz}.
a dependence of the electrical resistance in the normal on the magnetization
angle of the neighboring magnetic insulator, the spin Hall
magnetoresistance.\cite{Nakayama13,Chen13} We can therefore envisage a device
in which both spin valves are replaced by films of a metal with a large spin
Hall angle. In this case, the steady state magnetization angle is simply
$\theta=\arctan\left(  V_{B}/V_{S}\right)  $. However, since the spin Hall
magnetoresistance in the lower layer scales like $\alpha_{\mathrm{SH}}^{2}$,
such a device would be not very attractive unless $\alpha_{\mathrm{SH}}%
\gtrsim1,$ which has not been reported up to now. We therefore consider in the
following a hybrid device consisting of a source-drain lateral spin valve as
before, and only replace the upper one by a spin Hall metal.

We treat the upper layer (spin Hall system) by diffusion theory with quantum
mechanical boundary conditions at the interface to the insulating
magnet.\cite{Chen13} At $\theta=0\left(  \pi/2\right)  $ the source-drain
current into (or torque on) the magnetic insulator vanishes (is maximal)
while that from the upper film\ is maximal (vanishes). Following Ref.
\onlinecite{Chen13}%
, the torque reads:
\[
\tau_{B}^{\mathrm{SH}}(\theta)=\frac{\hbar}{2e}\alpha_{\mathrm{SH}}%
\frac{A\sigma}{L}V_{B}\frac{2\lambda G_{B2}^{\uparrow\downarrow}%
\tanh(d/2\lambda)}{\sigma+2\lambda G_{B2}^{\uparrow\downarrow}\coth
{(d/\lambda)}}\cos{\theta},
\]
where $d$ is the film thickness, $A=LW$ the cross section of the contact,
with $L$ and $W$ the length and width of the (rectangular) wire in contact
with the YIG disk, and $G_{B2}^{\uparrow\downarrow}(=e^2g_{B2}^{\uparrow\downarrow}/A/h)$ the real part of the spin-mixing interface conductance per unit area for the top contact.

\begin{figure}[ptb]
\begin{center}
\includegraphics[width=0.4\textwidth,angle=0]{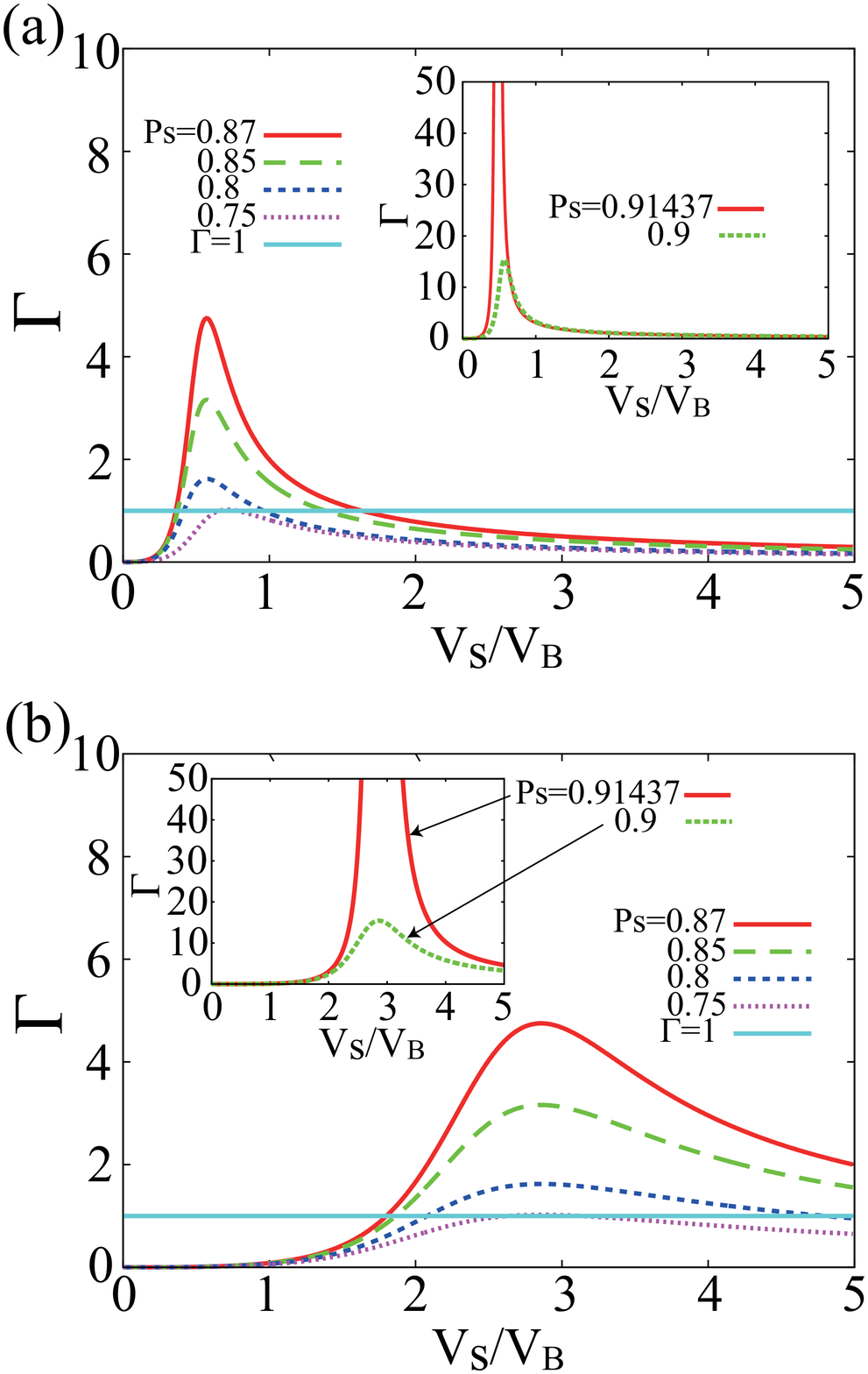}
\end{center}
\par
\caption{(a),(b):Differential current gain in the spin (Hall) transistor as a
function of the voltage ratio $V_{S}/V_{B}$ for different values of $p_{S}$
and $\alpha_{\mathrm{SH}}$ ($\beta=\gamma=5,e^{2}g_{S}/h/(L\sigma
)=1,K=5p_{S}/6/\alpha_{\mathrm{SH}}$) (c)$\alpha_{\mathrm{SH}}=0.3$, (d)
$\alpha_{\mathrm{SH}}=1.$ Insets represent the infinity gain for the critical value of $p_{S}$.}%
\label{Figure4}%
\end{figure}

The torque-induced rotation from $\theta=0$ suppresses the spin accumulation
and increases the source-drain current. As before this may lead to vanishing
differential conductance. We choose a model system with $p_{S}$ and
$\alpha_{\mathrm{SH}}$ variable, but other parameters fixed, viz. $g_{S}%
:g_{S}^{\uparrow\downarrow}:g_{B1}^{\uparrow\downarrow}:g_{B2}^{\uparrow
\downarrow}=1:1:\beta:\gamma$. The two torques on the base magnet cancel each
other, when $\tau_{B}^{N1}(\theta_{0})=\tau_{B}^{\mathrm{SH}}(\theta_{0})$ or
\begin{equation}
F\frac{V_{S}}{V_{B}}=\frac{\tan^{2}\theta_{0}+\epsilon}{\tan^{2}\theta_{0}%
+1}\frac{1}{\tan\theta_{0}}, \label{Eqltheta2}%
\end{equation}
where $F\equiv K(1-\epsilon),$ $\epsilon=\left(  \beta+2\right)
/(2\beta+2),$ and
\begin{equation}
K=2\frac{e^{2}g_{S}}{hA}\frac{L}{\sigma}\frac{p_{S}}{\alpha_{\mathrm{SH}}%
}\left[  \frac{\sigma}{2\gamma\lambda}\frac{hA}{e^{2}g_{S}}+\coth\frac{{d}%
}{{\lambda}}\right]  \coth\frac{{d}}{{2\lambda}}.
\end{equation}
The enhancement factor $F\sim L$ scales with the SHE metal wire and the base
magnetic insulator contact length because the spin current density is governed
by the electric field $V_{B}/L.$ The differential gain now reads%
\begin{widetext}
\begin{equation}
\Gamma=\frac{1}{F}\frac{2p_{S}^{2}\epsilon(1-\delta)\tan\theta}{\epsilon
+(3\epsilon-1)\tan^{2}\theta+\tan^{4}\theta-p_{S}^{2}\left[  \epsilon
+(3\epsilon-\delta)\tan^{2}\theta+\delta\tan^{4}\theta\right]  }. \label{DG22}%
\end{equation}
\end{widetext}
By substituting the solution of Eq. (\ref{Eqltheta2}), we calculate the differential
current gain $\Gamma$ as a function of $V_{S}/V_{B}\ $and plot it in Fig,
\ref{Figure4} for different values of the conductance polarization and the
spin Hall angle $\alpha_{\mathrm{SH}}$.

The critical value of $p_{S}$ at which $G(\theta)=0$ in Eq.
(\ref{DG22}) becomes
\begin{align}
p_{S}(\beta)  &  =\sqrt{\frac{9\epsilon-1}{9\epsilon-\delta}}=\sqrt
{\frac{7\beta+16}{9\beta+16}}\\
&  \rightarrow\frac{\sqrt{7}}{3}\simeq0.88192\ \ (\beta\rightarrow\infty).
\end{align}

The spin (Hall) torque transistors can display negative differential
resistance and differential gain by controlling source-drain current by the
competing spin transfers on both sides of the base magnetization. We
represented the negative differential resistance and gain as a function of the
ratio between the base voltage and the source-drain voltage. These device can
operate at room temperature, but in order to be useful, ferromagnetic
materials with polarizations close to unity and normal metals with a large
spin Hall angles are required. These parameters are still quite high, but
might be accessible with special materials. The base contact should be a
magnetic insulator in order to suppress undesired cross-talk and have a large
mixing conductance with the normal metal, which is known to be the case for
YIG.\cite{Jia11, Heinrich11} Tunnel junctions or reduced contact areas for the
source-drain contacts improve the differential gain but slow down the response
time and require reduced spin-flip scattering. The contact between the metals
and YIG should be relatively large. Since the current-induced torques due to
the spin Hall effect are comparable to that from spin valves, the performance
of the spin (Hall) torque transistor can be comparable to the old type, but
might be easier to fabricate. We also note that while $p_{S}\leq1$,
$\alpha_{\mathrm{SH}}$ is not limited by any principle.

The spin torque transistors are to our knowledge the only spintronics devices
that provide analogue gain; this in contrast to the Oersted field-operated
digital scheme in Ref.
\onlinecite{Konishi12}
(that, by the way, could also work with the spin Hall effect). A disadvantage
of the spin torque transistor is the stand-by current that is analogous to the
leakage current in bipolar transistors. The full electric control of the
magnetization direction without need for magnetic field might find
applications as well.

This work was supported by FOM (Stichting voor Fundamenteel Onderzoek der
Materie), EU-ICT-7 \textquotedblleft{MACALO,}\textquotedblright\ the ICC-IMR,
DFG Priority Programme 1538 \textquotedblleft{Spin-Caloric Transport}%
\textquotedblright\ (GO 944/4), and KAKENHI (No. 22540346).


\begin{thebibliography}{99}                                                                                               %


\bibitem {Datta90}S. Datta and B. Das, Appl. Phys. Lett. \textbf{56}, 665 (1990).

\bibitem {Konishi12}K. Konishi, T. Nozaki, H. Kubota, A. Fukushima, S. Yuasa,
and Y. Suzuki, IEEE Trans. Magn. \textbf{48}, 1134 (2012).

\bibitem {Bauer03}G. E. W. Bauer, A. Brataas, Y. Tserkovnyak, and B. J. van
Wees, Appl. Phys. Lett. \textbf{82}, 3928 (2003).

\bibitem {Uchida10}K. Uchida, J. Xiao, H. Adachi, J. Ohe, S. Takahashi, J.
Ieda, T. Ota, Y. Kajiwara, H. Umezawa, H. Kawai, G. E.W. Bauer, S. Maekawa,
and E. Saitoh, Nature Mater. \textbf{9}, 894 (2010).

\bibitem {Kajiwara10}Y. Kajiwara, K. Harii, S. Takahashi, J. Ohe, K. Uchida,
M. Mizuguchi, H. Umezawa, H. Kawai, K. Ando, K. Takanashi, S. Maekawa, and E.
Saitoh, Nature \textbf{464}, 262 (2010).

\bibitem {Brataas00}A. Brataas, Yu. V. Nazarov, and G. E. W. Bauer, Phys. Rev.
Lett. \textbf{84}, 2481 (2000); Eur. Phys. J. B \textbf{22}, 99 (2001).

\bibitem {Jia11}X. Jia, K. Liu, K. Xia, and G. E. W. Bauer, Europhys. Lett.
\textbf{96}, 17005 (2011).

\bibitem {Heinrich11}C. Burrowes, B. Heinrich, B. Kardasz, E. A. Montoya, E.
Girt, Y. Sun, Y. Y. Song, and M. Wu, Appl. Phys. Lett. \textbf{100}, 092403 (2012).

\bibitem {Tant}L. Liu, C.-F. Pai, Y. Li, H. W. Tseng, D. C. Ralph, R. A.
Buhrman, Science \textbf{336}, 555 (2012).

\bibitem {Tung}C.-F. Pai, L. Liu, Y. Li, H. W. Tseng, D. C. Ralph, and R. A.
Buhrman, Appl. Phys. Lett. \textbf{101}, 122404 (2012).

\bibitem {CuBi}Y. Niimi, Y. Kawanishi, D. H. Wei1, C. Deranlot, H. X. Yang, M.
Chshiev, T. Valet, A. Fert, and Y. Otani, Phys. Rev. Lett. \textbf{109},
156602 (2012).

\bibitem {Xia10}K. Xia, P. J. Kelly, G. E. W. Bauer, A. Brataas, and I. Turek,
Phys. Rev. B \textbf{65}, 220401 (2002).

\bibitem {Stiles12}M. D. Stiles and A. Zangwill, Phys. Rev. B \textbf{66},
14407 (2002).

\bibitem {Jedema00}F. J. Jedema, H. B. Heersche, A. T. Filip,
J. J. A. Baselmans, and B. J. van Wees, ibid. \textbf{416}, 713 (2002); S. O. Valenzuela and M. Tinkham, Nature \textbf{442}, 176 (2006).

\bibitem {Jedema01}F. J. Jedema, A. T. Filip, and B. J. van Wees, Nature
(London) \textbf{410}, 345 (2000).

\bibitem {Silver}T. Kimura and Y. Otani, Phys. Rev. Lett. \textbf{99}, 196604 (2007).

\bibitem {graphene}M. Wojtaszek, I. J. Vera-Marun, T. Maassen, and B. J. van
Wees, Phys. Rev. B \textbf{87}, 081402(R) (2013)

\bibitem {Bauer10}G. E. W. Bauer, Y. Tserkovnyak, D. Huertas, and A. Brataas,
Phys. Rev. B \textbf{67}, 094421 (2003).

\bibitem {SA}A. Brataas, Y.V. Nazarov, J. Inoue, G.E.W. Bauer, Phys. Rev. B
59, 93 (1999); Eur. Phys. J. B 9, 421 (1999).

\bibitem {Jedema03}F. J. Jedema, M. S. Nijboer, A. T. Filip, and B. J. van
Wees, Phys. Rev. B \textbf{67}, 085319 (2003); J. Bass and W. P. Pratt. J.
Phys. Condens. Matter \textbf{19}, 183201 (2007).

\bibitem {Han12}W. Han and R. K. Kawakami, Phys. Rev. Lett. \textbf{107},
047207 (2011); M. H. D. Guimar\~{a}es, A. Veligura, P. J. Zomer, T.
Maassen, I. J. Vera-Marun, N. Tombros, and B. J. van Wees, Nano Lett.
\textbf{12}, 3512 (2012).

\bibitem {Dyakonov10}M. I. Dyakonov and V. I. Perel, Phys. Lett. A
\textbf{35}, 459 (1971); J. E. Hirsch, Phys. Rev. Lett. \textbf{83}, 1834
(1999).;S. F. Zhang, Phys. Rev. Lett. \textbf{85}, 393 (2000).

\bibitem {Kimura10}T. Kimura, Y. Otani, T. Sato, S. Takahashi, and S. Maekawa,
Phys. Rev. Lett. \textbf{98}, 156601 (2007); K. Ando, S. Takahashi, K. Harii,
K. Sasage, J. Ieda, S. Maekawa, and E. Saitoh, Phys. Rev. Lett. \textbf{101},
036601 (2008); O. Mosendz, J. E. Pearson, F. Y. Fradin, G. E. W. Bauer, S. D.
Bader, and A. Hoffmann, Phys. Rev. Lett. \textbf{104}, 046601 (2010); L. Q.
Liu, T. Moriyama, D. C. Ralph, and R. A. Buhrman, Phys. Rev. Lett.
\textbf{106}, 036601 (2011); L. Q. Liu, O. J. Lee, T. D. Gudmundsen, R. A.
Buhrman, and D. C. Ralph, arXiv:1110.6846; L. Q. Liu, R. A. Buhrman, and D. C.
Ralph, arXiv:1111.3702.

\bibitem {Nakayama13}H. Nakayama, M. Althammer, Y.-T. Chen, K. Uchida, Y.
Kajiwara, D. Kikuchi, T. Ohtani, S. Gepr\"{a}gs, M. Opel, S. Takahashi, R.
Gross, G. E. W. Bauer, S. T. B. Goennenwein, E. Saitoh, arXiv:1211.0098..

\bibitem {Chen13}Y. Chen, S. Takahashi, H. Nakayama, M. Althammer, S. T. B.
Goennenwein, E. Saitoh, G. E. W. Bauer, arXiv:1302.1352.
\end{thebibliography}
\end{document}